\documentclass[aps,prb,reprint,superscriptaddress,floatfix,showpacs]{revtex4-1}
\usepackage{dcolumn}
\usepackage{bm}
\usepackage[utf8]{inputenc}
\usepackage{color}
\usepackage{hyperref}
\usepackage{siunitx}
\usepackage{amsmath}
\usepackage{bbm}
\usepackage{ulem}
\usepackage{float}
\definecolor{purple}{RGB}{128,0,128}
\definecolor{orange}{RGB}{255,135,0}

\newcommand{\de}{\ensuremath{^{\circ}}}
\renewcommand{\vec}[1]{\bm{#1}}

\usepackage{graphicx}
\graphicspath{{./Figures/}}

\begin{document}
	
\title{Giant atomic swirl in graphene bilayers with biaxial heterostrain.}


	\author{Florie Mesple}
	
	\affiliation{Univ. Grenoble Alpes, CEA, Grenoble INP, IRIG, PHELIQS, 38000 Grenoble, France}
	\affiliation{Department of Physics, University of Washington, Seattle, Washington, 98195, USA}
	
	\author{Niels R. Walet}
	\affiliation{Department of Physics and Astronomy, University of Manchester, Manchester, M13 9PY, United Kingdom}
	
	\author{Guy Trambly de Laissardière}
	\affiliation{Laboratoire de Physique Théorique et Modélisation (UMR 8089), CY Cergy Paris Université, CNRS, 95302 Cergy-Pontoise, France}

	\author{Francisco Guinea}
	\affiliation{Imdea Nanoscience, Faraday 9, 28015 Madrid, Spain}
	\affiliation{Donostia International Physics Center, Paseo Manuel de Lardizábal 4, 20018 San Sebastián, Spain}

	\author{Djordje Do\v{s}enovi\'{c}}
	\author{Hanako Okuno}
	\affiliation{University Grenoble Alpes, CEA, IRIG-MEM, Grenoble, France}
	
	\author{Colin Paillet}
	\author{Adrien Michon}
	\affiliation{Université Côte d’Azur, CNRS, CRHEA, Rue Bernard Grégory, 06560 Valbonne, France}

	\author{Claude Chapelier}
	\affiliation{Univ. Grenoble Alpes, CEA, Grenoble INP, IRIG, PHELIQS, 38000 Grenoble, France}
	\author{Vincent T. Renard}
	\email{email: vincent.renard@cea.fr}
	\affiliation{Univ. Grenoble Alpes, CEA, Grenoble INP, IRIG, PHELIQS, 38000 Grenoble, France}

	\date{August 25, 2023}
	
	

	\begin{abstract}
The study of moiré engineering started with the advent of van der Waals heterostructures in which stacking two-dimensional layers with different lattice constants leads to a moiré pattern controlling their electronic properties. The field entered a new era when it was found that adjusting the twist between two graphene layers led to strongly-correlated-electron physics and topological effects associated with atomic relaxation. Twist is now used routinely to adjust the properties of two-dimensional materials. Here, we investigate a new type of moiré superlattice in bilayer graphene when one layer is biaxially strained with respect to the other - so-called biaxial heterostrain. Scanning tunneling microscopy measurements uncover spiraling electronic states associated with a novel symmetry-breaking atomic reconstruction at small biaxial heterostrain. Atomistic calculations using experimental parameters as inputs reveal that a giant atomic swirl forms around regions of aligned stacking to reduce the mechanical energy of the bilayer. Tight-binding calculations performed on the relaxed structure show that the observed electronic states decorate spiraling domain wall solitons as required by topology. This study establishes biaxial heterostrain as an important parameter to be harnessed for the next step of moiré engineering in van der Waals multilayers.
	 
	\end{abstract}

\maketitle

\begin{figure*}
	\centering
	\includegraphics[width=0.9\linewidth]{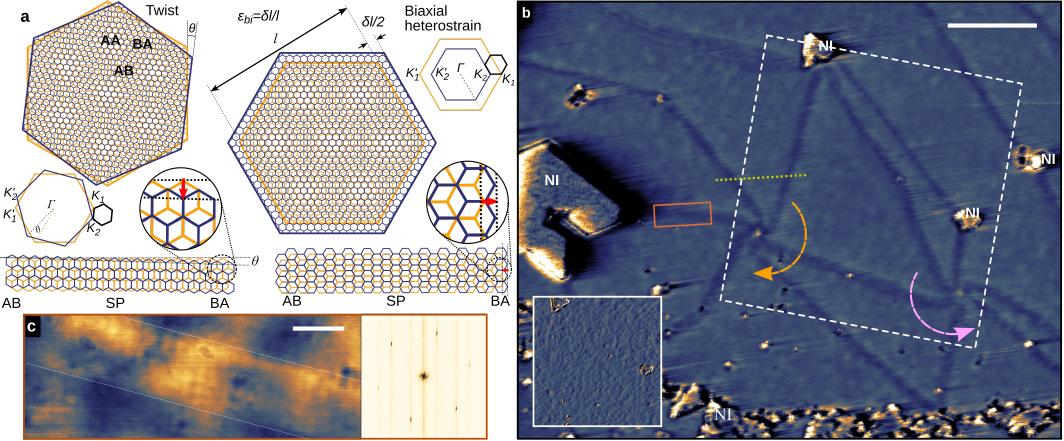}
	\caption{\textbf{Soliton network in bilayer graphene on SiC} \textbf{a} Schematic model of the real and reciprocal space structure of a twist moiré (left) and biaxial moiré (right) with their corresponding domain walls - of shear and tensile types, respectively. The Burgers vector (in red) characterises the DW type as it is either parallel to the DW for a shear DW or perpendicular to it for a tensile one. \textbf{b} STM current image ($V_b = -300\,\text{mV}$, $I_t = 250\,\text{pA}$) for an intercalated surface of bilayer graphene on SiC, with a reconstructed moiré featuring a (anti-)clockwise swirling feature in (blue) orange. Non-intercalated regions are indicated by NI. The scale bar is 100 nm long. The inset shows the STM image of the area highlighted by the dotted square measured at $V_b = 900\,\text{mV}$ and $I_t = 250\,\text{pA}$. The dotted yellow line shows the position of the line spectroscopy presented in Fig.~\ref{fig:PropElec}f. \textbf{c} Enlarged view of a soliton (red rectangle in b and for the same tunneling conditions). The Fourier transform shows that the crystallographic directions (rotated by $\pi/2$ with respect to Fourier harmonics) are along the soliton direction. The scale bar is 10 nm long.}
	\label{fig:SiC}
\end{figure*}

The physics of moiré materials is driven by the details of the relative arrangement of two-dimensional layers stacked on top of each other. The emblematic example is twisted bilayer graphene (TBG) which evolves from two independent layers for rotation angles larger than $10^\circ$\cite{LopesdosSantos2007,Li2009} to a flat-band material showing strongly correlated electron physics near the magic angle of $1.1^\circ$.\cite{Cao2019,Cao2019b} For angles smaller than the magic one, domain wall solitons with topological one-dimensional electronic states\cite{san-jose_helical_2013,zhang_valley_2013,Rickhaus2018} form to minimize the stacking energy\cite{Wijk_2015,Alden2013,Walet_2020}. This rich physics is determined by a single parameter: the twist angle between the layers which can be controlled precisely in the tear and stack technique.\cite{Tutuc2016} Yet, it is also possible to change the relative arrangement of graphene layers using heterostrain, the deformation of one layer with respect to the other.\cite{Huder2018} Until now research has largely focused on the study of non-intentional uniaxial heterostrain which broadens the flat bands\cite{Bi2019designing,mesple_heterostrain_2021} and is essential in the selection of the many-body ground states of TBG.\cite{Nuckolls2023} Much less attention has been devoted to biaxial heterostrain which is generally of smaller amplitude and was assumed to have a smaller effect on the physics of TBG.\cite{nakatsuji_moire_2022}  However, bilayer graphene with pure biaxial heterostrain presents a moiré pattern with Bernal stacked (AB) and aligned (AA) regions, similar to TBG (Fig.~\ref{fig:SiC}a). This moiré has 6-fold symmetry but, contrary to that of TBG, it is aligned with graphene's crystallographic directions.\cite{Van_der_Donck_2016} Figure~\ref{fig:SiC}a also reveals that the domain walls (DWs) connecting Bernal stacked AB and BA regions are tensile while those of TBGs are shear. This is quantified by the orientation of the Burgers vector describing the shift of the unit cell across the DW (Fig.~\ref{fig:SiC}a).\cite{vaezi_topological_2013} These structural differences raise the question of the electronic properties of such biaxial heterostrain moiré and whether or not they can be observed experimentally.

\subsection*{Morphology of biaxially-heterostrained graphene bilayer.}
Figure~\ref{fig:SiC}b shows a scanning-tunnelling-microscope (STM) image of bilayer graphene produced by hydrogen intercalation of graphene on the silicon face of 6H-SiC \cite{Michon2010, Michon2018, Joualt2014, riedl_quasi-free-standing_2009} (See Methods for details). The surface is mostly released although a few non-intercalated (NI) regions remain (Fig.~\ref{fig:SiC}b and 
Supplements). The intercalated surface features a complex network of lines similar to the soliton network observed in marginally-twisted bilayer graphene (MTBG). However, here the solitons spiral around their crossing points which has not been reported in MTBG. \cite{Alden2013,Yoo2019,halbertal_moire_2021,de_jong_imaging_2022}
In addition, atomically resolved images (Fig.~\ref{fig:SiC}c) show that the solitons are mostly aligned with graphene's crystallographic directions suggesting that the layers are subject to biaxial heterostrain.
The soliton length of about $a_\text{M}=250\,\text{nm}$ gives an estimate of $\varepsilon_\text{bi}=a_\text{Gr}/a_\text{M}\approx0.1\%$ where $a_\text{Gr}$ is the graphene lattice constant.
This value is consistent with previous measurements\cite{michon_graphene_2012,giannazzo_probing_2019,de_jong_stacking_2023} which concluded that strain is imposed to the bottom layer by the $6\sqrt{3}\times 6\sqrt{3}R30$\de SiC reconstruction.

\subsection*{Modeling of the structural relaxation.}
Previous investigations of monolayers of metallic atoms on a rigid substrate show that a small lattice mismatch can lead to such spiralling solitons\cite{Barth1994,gunther_strain_1995,corso_au111-based_2010,quan_tunable_2018}.
Before we demonstrate that this also the case for bilayer graphene, a more detailed investigation of the relative arrangement of the layers is needed. The solitons in Fig.~\ref{fig:SiC}b do not all have the same length suggesting a uniaxial heterostrain component which varies in space.
It is therefore difficult to describe the entire structure directly. Instead we model each of the triangles as a distinct commensurate structure (See supplements). 
We focus on the one highlighted by the dotted square in Fig.~\ref{fig:SiC}b which is characterized by a biaxial heterostrain of $\varepsilon_\text{bi}=$-0.08\%, a uniaxial heterostrain of $\varepsilon_\text{uni}=$-0.057\% and a twist angle of $\theta=0.01^\circ$. We then build a 3.5-million-atoms commensurate cell with this particular relative stacking describing a situation with homogeneous strain.\cite{mesple_heterostrain_2021}
However, this is still not realistic because the layers relax to balance in-layer strain and interlayer atomic alignment in such a huge moiré lattice. We therefore perform an atomistic simulation minimizing the stacking and elastic energies, as detailed in the Methods section and Ref.~\onlinecite{Walet_2020}. 

\begin{figure*}[ht]
	\centering
	\includegraphics[width=0.93\linewidth]{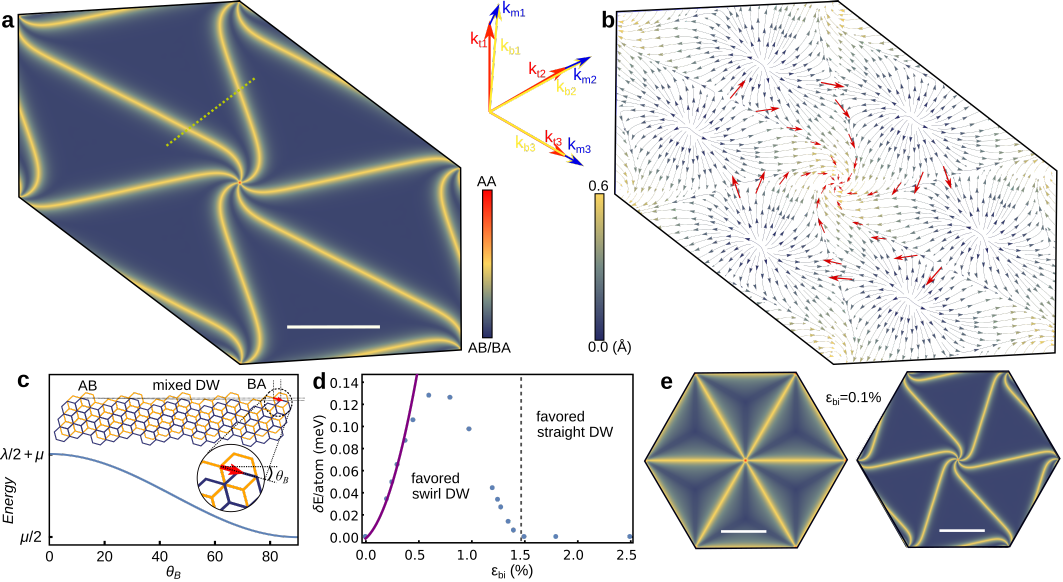}
	\caption{\textbf{Structural relaxation}. \textbf{a} Calculated stacking parameter: blue shows AB/BA alignment, red AA, and yellow shows competing alignment. (See Ref.~\onlinecite{guinea_continuum_2019} for the definition of the stacking parameter.) The dotted line shows the position of the LDOS calculations of Fig.~\ref{fig:PropElec}g. The scale bar is 100nm. \textbf{b} Atomic-flow map leading to the relaxed structure. The arrows show qualitative streamlines of one layer where lighter colors indicate larger atomic movements. In the other layer, the atoms move in the opposite direction. The red arrows indicate the local Burgers vectors along the domain walls. \textbf{c} Relative stacking energy of the soliton upon variation of the soliton type, characterized by the angle between the Burgers vector and the direction perpendicular to the DW, as defined on the sketch of a DW of mixed type. \textbf{d} Energetics of configurations with varying biaxial heterostrain: We show the difference between the swirl and straight solitons relaxation modes. Above $\varepsilon_\text{bi}=1.4\%$ (vertical dotted line), no swirl solution can be found (See supplements). The purple line shows a simple fit for small heterostrain,  and is valid only up to $\sim 0.5\%$ strain (Supplements). \textbf{e} Stacking parameter for a relaxed structure from a moiré induced by a $0.1$\% purely biaxial heterostrain. Left shows the low energy swirl soliton mode, and right the metastable straight soliton mode. The scale bars are 100nm.
	}
	\label{fig:Relaxation}
\end{figure*}

The resulting local stacking represented in colors in Figure \ref{fig:Relaxation}a reproduces very well spiralling solitons. In principle, the relaxation can feature either spiraling or straight solitons for the same parameters (See Fig.~\ref{fig:Relaxation}e for a 0.1\% purely biaxial moiré). But the spiral relaxation mode dominates up to about $1.5\%$ of strain (Fig.~\ref{fig:Relaxation}d). Similarly to marginally-twisted graphene layers, AB regions which have lower stacking energy grow at the expense of the higher-energy AA regions, concentrating elastic strain in the solitons.\cite{Wijk2015} Interestingly, while AB regions are marginally affected by it, the spiral mode drastically reduces the surface of AA regions bellow $\varepsilon_{bi}=$1.5\% (supplements). In addition, atoms flow outward from the center of the AB regions in the top layer forming a giant atomic swirl around AA regions (Fig.~\ref{fig:Relaxation}b) further reducing the elastic energy. Indeed, the swirl causes the Burger vector of the soliton (defined in Fig.~\ref{fig:Relaxation}c) to rotate from perpendicular in between AA regions to parallel in AA region, which reduces its energy according to the following equation:
\begin{equation}
E_\text{elas} \propto
\left[  \left(\frac{\lambda}{2}+\mu \right) \cos ^ {2} (\theta_B) + \mu \frac{\sin ^ {2} (\theta_B)}{2} \right] 
\label{Eq: Elastic energy}
\end{equation}
where $\lambda$ and $\mu$ are the first and second Lamé parameters of graphene and $\theta_B$ the angle of the Burger vector (Fig~\ref{fig:Relaxation}c and Methods for more details). Even though derived by elastic theory, this reflects the behaviour at the atomic scale: in general it is harder to change bond lengths than bond angles. Supplements shows that the solitons fully relax into shear type in AA region.

In addition, calculations of  Figs.~\ref{fig:Relaxation}a and e show that the spiral is mostly determined by the biaxial heterostrain component while uniaxial heterostrain has a small effect. Relaxation restores locally the rotational symmetry near AA regions which is lost at the scale of the unit cell due to the uniaxial heterostrain. This explains the regularity of the spiral seen in experiment despite the in-homogeneity of heterostrain and justifies our approximation of the experimental structure by the chosen unit cell. The experimental diameter of the swirl (about 100 nm) and the width of the soliton, between 15 nm and 35 nm depending on energy (see below), are very well reproduced ($20\pm10~\text{nm}$ in the theory).  Supplements presents complementary calculations of the development of the swirl as function of biaxial heterostrain. Figure~\ref{fig:SiC}b suggests that there exist no preferred direction for the swirl. In numerical simulations we find both orientations depending on some random inputs that give rise to the symmetry breaking. This is clearly evidenced in he supplements, where we show the reaction path from a left- to a right-rotating swirl, with the straight DW case as the transition state between the two, i.e., at a saddle point in the energy landscape.  It is thus a classic example of spontaneous symmetry breaking through a pitch-fork bifurcation, caused by the applied heterostrain.\cite{Engelke_2022,cazeaux_relaxation_2022} As a final verification we checked that no swirl is obtained in the calculations of a twist moiré (Supplement). All in all, the theoretical results together with the interpretation of the experimental images firmly establish that we have indeed observed a marginally biaxially heterostrained graphene bilayer.  

\begin{figure*}[t]
	\centering
	\includegraphics[width=0.95\linewidth]{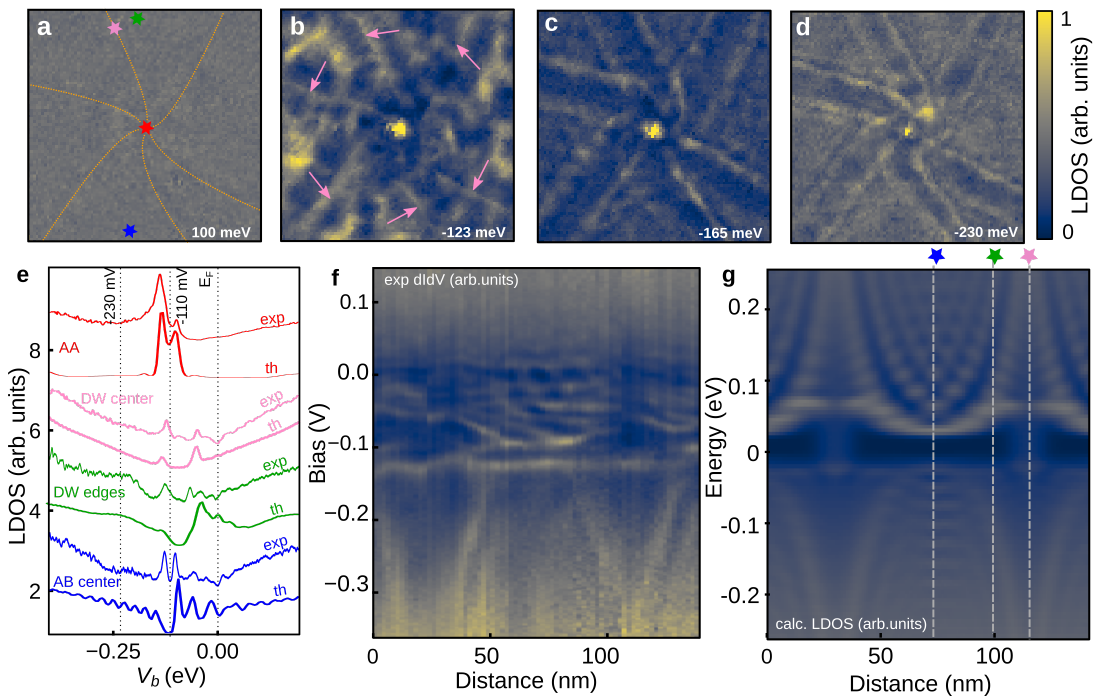}
	\caption{\textbf{Electronic properties of a marginally biaxially heterostrained graphene bilayer}. \textbf{a-d} Spatially-resolved LDOS maps centred around the AA region, featuring energy-dependent spectroscopic features. The setpoint is ($V_{b} = +200~\text{mV}$, $I_{t} = 250~\text{pA}$)  \textbf{e} Corresponding individual spectra showing localized states in the AA region in red, a localized state at the center of the DW at charge neutrality in pink, and energy dependent confined states at the DW edges (green, orange). The Dirac energy $E_D$ is shifted with respect to $E_F$ due to the substrate induced doping. The electric field induced gap in AB regions allows to define charge neutrality at $-110$ meV. Each experimental spectrum is shown along its tight-binding-calculated counterpart, which describe the data well. \textbf{f} Experimental LDOS along the yellow dotted line from Fig.~\ref{fig:SiC}b that crosses two solitons. \textbf{g} Corresponding tight-binding calculations, taken along the yellow dotted line defined in Fig.~\ref{fig:Relaxation}a. }
	\label{fig:PropElec}
\end{figure*}

\subsection*{Solitonic electronic states}

Interestingly, while the soliton network is visible at negative energies, the STM image is essentially featureless at positive energy (Fig.~\ref{fig:SiC}b,c and supplements). This strongly suggests that solitonic features seen in STM are of electronic origin calling for a detailed study. Figure~\ref{fig:PropElec}a-d show local-density-of-states (LDOS) maps determined from a differential conductance spectrum measured at each pixel (See Methods and supplementary animation for the full energy dependence of the maps). The measurements are centred on the central spiral in Fig.~\ref{fig:SiC}b. The sequence of energy-resolved images reveal a complex spatial dependence of the spectroscopic features. 
First of all, AB regions far from the solitons show two resonances which we attribute to an electronic gap opened in AB region by the electric field induced by the substrate (Fig.~\ref{fig:PropElec}e, blue). This allows us to determine a gap of $E_g=27$ meV and charge neutrality point at $E_D=-110$ meV, both of which are typical of bilayer graphene on SiC\cite{rutter_imaging_2007,lauffer_atomic,botswick2009,moreau_high-resolution_2013}. We also note a V-shaped depression of the experimental density of states at the Fermi energy $E_F$ which we attribute to the tunnelling anomaly in two-dimensional systems\cite{Altshuler1980} (See also Fig.~\ref{fig:PropElec}f). A strong pair of peaks indicative of resonance is found at $E_D$ in the center of the swirl. The corresponding spectrum shown in Fig.~\ref{fig:PropElec}b (red) strongly resembles the localization peak seen in AA regions of TBG. \cite{Li2009,Trambly2010}
Below charge neutrality, we observe a set of edge-state-like (ESL) features flanking the solitons and moving towards it as energy is decreased Fig.~\ref{fig:PropElec}c-d. These lead to an energy-dependent effective width of the soliton seen in STM images and correspond to broad resonances in individual spectra (Fig.~\ref{fig:PropElec}e, green). Similar features were already reported in TBG and attributed to either van Hove singularities \cite{Huang_2018}, or to pseudo Landau Levels induced by the inhomogeneous strain concentrated at the DW.\cite{zheng_tunable_2022} We exclude these interpretations here because of the broad spatial extension and highly dispersive nature of the ESL. 

In order to get more insight into these spectral features, we perform tight-binding calculations using the relaxed unit cell shown in Fig.~\ref{fig:Relaxation}a, including a 30 meV potential difference between the layers as in the experiment (See Methods and Ref.~\onlinecite{Trambly2010} for details on TB calculations). The local density of states calculated in the top layer reproduces well the experimental one (Fig.~\ref{fig:PropElec}e). However they do not shed light on the origin of ESL features, and we show the full spatial dependence along a line crossing the unit cell in Fig.~\ref{fig:PropElec}f (experiments) and Fig.~\ref{fig:PropElec}g (theory). 
Away from $E_F$, the local density of states reveals confined states in the AB regions which express more intensely above charge neutrality, 
and are reminiscent of a 2D electron gas confined by hard walls. \cite{veuillen_electron_2003} The confinement originates from the reflection of electrons by the tensile domain wall which occurs up to high energies at normal incidence for tensile solitons.\cite{Koshino_2013, san-jose_stacking_2014} It is interesting that the present soliton has a rotating Burgers vector which does not seem to affect the confinement and will deserve further theoretical attention.  Another interesting feature is that 
below $E_D$, only the first interference oscillations decorate the confinement regions and disperse like ESL features in experiments. This particle-hole asymmetry is reproduced in the calculations. At $E>E_D$ the localized states manifest as sharp resonances distributed on the entire AB region with a characteristic length scale (Fig.~S\ref{fig:Suppl:Triangle} and supplementary animation for the full energy dependence of the maps) which are not perfectly periodic because of the irregularity of the confining region and fluctuations of the Fermi level.
Noteworthy, the LDOS is featureless above $E_F$ as expected from STM images.  

Finally we address the question of the states located in the soliton's core. These are emphasized in Fig.~\ref{fig:PropElec}b and e (pink arrows and LDOS spectrums) and well captured in our tight-binding calculations that include both heterostrain and relaxation. They are slightly obscured in the experiment due to the strong confinement features in the neighbouring AB regions at the same energies.
Topology requires the presence of helical states in such channels created by the solitons at the AB/BA boundary.\cite{san-jose_helical_2013,zhang_valley_2013,Rickhaus2018}
It is therefore tempting to attribute these core states to helical states. However, helical states which are not topologically protected could possibly hybridize with the bulk states.\cite{Walet_2020} Moreover, it is not clear what is the impact of the pseudomagnetic field that arises due to the strain gradient and which will certainly have a peculiar texture in the spiraling soliton.
This will deserve more theoretical investigations with a phenomenological model including both hybridization and pseudomagnetic field. 

\subsection*{Perspectives}

Biaxially heterostrain offers a new moiré platform to study localization and topological solitons. Its interplay with other parameters such as twist angle and large uniaxial heterostrain should be investigated more systematically. Also, even if our study demonstrates that in-plane atomic movement is the main driver of our observation, out of plane relaxation will have to be considered. This will require a much more involved, and not so well established, tight-binding model\cite{guinea_continuum_2019} (See also  Ref.~\onlinecite{dai_twisted_2016,rakib_corrugation-driven_2022} for related works in TBG). 
From the experimental perspective, we have exploited the native biaxial heterostrain in intercalated bilayer graphene on SiC which is certainly stabilized by non intercalated regions. Experiments in trilayer graphene\cite{lalmi_flower_shaped_2015} and WS$_2$\cite{li_morphology_2022} with similar morphology indicate that the atomic swirl also occurs in other van de Waals stacks under appropriate conditions and this is supported by very recent calculations\cite{kaliteevsky2023twirling} in MoX$_2$ /WX$_2$ (TMDs with X=S or Se). A more systematic study will require to control biaxial heterostrain as was recently demonstrated in Ref.~\onlinecite{pasquier_tunable_2022}. We anticipate that the combined choice of material, twist angle, heterostrain and pressure will bring moiré engineering to its full potential.

\vspace{50pt}

\section*{Methods}

\subsection*{Sample fabrication}
The sample was grown by propane-hydrogen Chemical Vapour Deposition (CVD) on the Si-face of 6H-SiC.\cite{Michon2010} During the whole process, SiC is exposed to a 91\%Ar 9\%H2 mixture at a pressure of 800 mbar. After a 5 minutes temperature ramp, propane is added to the gas phase for a 15 minutes growth plateau at 1550\de C, and finally, the sample is cooled to room temperature without propane within 20 minutes. This process can be compared to the Si-sublimation method. However, under hydrogen, carbon from SiC cannot be segregated on the surface, making an external carbon source mandatory to grow graphene.\cite{Michon2018} Nevertheless, for the growth conditions specified above, the graphene film is very similar to what can be obtained using Si-sublimation, i.e., a monolayer graphene resting on a buffer layer on SiC.\cite{Joualt2014} This buffer layer is made of carbon atoms arranged in an honeycomb lattice similar to graphene, but containing a significant fraction of $sp^3$ carbon atoms covalently bonded to SiC. These bonds can be detached by intercalating Hydrogen atoms below the buffer layer, which  produces a quasi-free-standing bilayer graphene.\cite{riedl_quasi-free-standing_2009}
In this work the intercalation is done by a 30 minutes annealing under a H$_2$/NH$_3$ gas mixture at 150 mbar and at a temperature of 1100\de C. 

\subsection*{Scanning Tunnelling Microscopy and Spectroscopy}
The measurements are acquired using a homebuilt STM in a cryogenic environment at $4.2\,\text{K}$ that ensures a $\sim 1.5\,\text{meV}$ resolution on spectroscopic measurements. We identify each measurement with its setpoint: the bias voltage $V_b$ and tunnelling current $I_t$ at the start of the bias sweep.
The spectroscopic measurements were recorded using phase sensitive detection with a 4mV AC voltage modulation. The density of state maps are obtained from grid spectroscopy - or Current-Imaging Tunnelling Spectroscopy (CITS). At each pixel, the feedback loop is switched off at the setpoint conditions and a sweep over the bias voltage is performed  while recording the $dI/dV$ signal. All microscopic images and CITS data are analysed using Gwyddion \cite{Gwyddion} and STMaPy (Scanning Tunneling Microscopy analysis in PYthon), a  data analysis tool written in Python by some of the authors\cite{huder_loic_2023_7971228}.

\subsection*{Local commensurably analysis of the moiré swirl pattern}

Our STM images reveals a moiré with a very large periodicity, thus associated with atomic relaxation and atomic reconstruction phenomena.
Nevertheless, the dimensions of the moiré can be associated with the corresponding unrelaxed stacking. 
Following Refs.~\onlinecite{Artaud2016,Huder2018} we describe the relative stacking with the twist $\theta_\text{int} = \theta$ between the layers, uniaxial $\varepsilon_\text{uni}$ and biaxial $\varepsilon_\text{bi}$ heterostrain.
The analysis of the moiré length has to be combined with an analysis of the local crystallographic directions of graphene in order to find a unique stacking. These can be determined exactly from atomically resolved STM images, which is spatially homogeneous, with variations of 0.5\de at the DWs where strain is concentrated by the reconstruction.

In practice, we solve the following equation :

\begin{align}
\begin{pmatrix} \vec{k}_{b_1} \\ \vec{k}_{b_2} \end{pmatrix} = \begin{pmatrix} a & c \\b & d \end{pmatrix} \begin{pmatrix} \vec{k}_{t_1} \\ \vec{k}_{t_2} \end{pmatrix}
\label{eq:ParkMaddenk}
\end{align}

where the top layer periodicity in reciprocal space $\vec{k}_{t_i}$, $i=1,2$ is determined from atomically resolved STM images, and the bottom layer periodicity $\vec{k}_{b_i}$ is determined from the first order commensurably equation $\vec{k}_{b_i} = \vec{k}_{t_i} + \vec{k}_{m_i}$ where $\vec{k}_{m_i}$ is the moiré period of a given triangle determined from large scale images.

This relative stacking is described by the so called Park-Madden matrix as introduced in Ref.~\onlinecite{Artaud2016} and used in refs.~\onlinecite{Huder2018,mesple_heterostrain_2021} describes the relative stacking of the layers and experimental $(a, b, c, d)$ can be related to the rotation between the layers and heterostrain. The moiré size is set by the values of pure twist and biaxial heterostrain between the layers while the anisotropy of the moiré is characterized by the uniaxial heterostrain. Note that such big moiré lengths are very sensitive to the relative stacking variables, and so we find those variables by imposing the exact moiré size required.

Because of the high inhomogeneity of the moiré, we perform this analysis for each closed triangle of the moiré as shown on Fig.~S2,
where the relative stacking including twist and heterostrain are attributed to each individual triangle. 

Note that this analysis describes the unrelaxed stacking of the moiré, which is locally modified upon relaxation, concentrating strain at the DWs and around the AA-stacking region and maximizing the surface area of AB and BA stacked regions.

\subsection*{Calculations of relaxation}

From the Park-Madden matrix as derived from previous section, we use the method presented in Ref.~\onlinecite{mesple_heterostrain_2021} to produce a commensurate cell describing the unrelaxed relative stacking between the layers.
We then find the lowest energy relaxed configuration using molecular dynamic calculations using LAMMPS.\cite{LAMMPS}
These calculations minimize binding and elastic energy of the layers by allowing in-plane movements of atoms. The method and ideas specific to graphene are discussed in Ref.~\onlinecite{guinea_continuum_2019}, as well as references cited therein. We use the REBO (Reactive Empirical Bond Order) potential to describe intralayer interactions \cite{REBO}, and the Kolmogorov-Crespi \cite{KC} potential to describe the interlayer interactions. 
Even though the global effect of the relaxation are large, the most an individual atom is displaced in plane is $0.6~\text{\AA}$. More details on these displacements are shown in Extendend Data Fig.~S\ref{fig:Suppl:MoirePars}.

\subsection*{Tight-binding}

All tight-binding calculations in this paper use parameters that have been determined previously \cite{Brihuega2012} and show good agreement with experimental data.\cite{Huder2018,mesple_heterostrain_2021} 
With these parameters, the Fermi velocity in a monolayer is $v_\text{mono} = 1.1\times 10^6 \text{m/s}$.
In AB/BA bilayers, the sublattice asymmetry is significant in the LDOS spectra for energies below $450~\text{meV}$ as the lowest energy dispersive bands are localized on the non-dimer atoms of the layers ($A_1/B_2$ or $A_2/B_1$) ; See Fig.~S\ref{fig:Suppl:TB} for more insight.
The LDOS calculations shown in this work correspond the averaged LDOS between the A and B sublattices, using a $5~\text{meV}$ Gaussian broadening.
In addition, a $30~\text{meV}$ potential difference between the layers is included, to correspond to the experimental situation.

\subsection*{Energy of the soliton}

The relaxation calculations show that the giant swirl orignates from a rotation of the Burgers vectors of the DW as the soliton varies from a tensile type to a shear type.
Normally, a shear type soliton is energetically favoured, as it implies requires a smaller modification of the carbon-carbon bond lengths.

This can be modelled by a continuous local displacement $\bm{u}$ of the atoms along a soliton, expressed in terms of the displacement along the soliton as given by the function $f(x)$, modulated by the direction of their displacement as defined by $\theta_B$, the angle of the Burgers vector with respect to the DW direction.
This angle characterises the soliton type, which ranges from purely tensile ($\theta_B = 0$) to purely shear ($\theta_B = \pi/2$).
\begin{equation}
\left( u_{x}(x) , u_{y} (x)\right) = \left( f (x) \cos (\theta_B), f(x) \sin(\theta_B) \right)\,,
\end{equation}
where the total displacement across the soliton dislocation satisfies
\begin{equation}
f(x) = 
\begin{cases}
- \frac{d}{2 \sqrt{3}} & x\rightarrow -\infty \\
\frac{d}{2 \sqrt{3}} & x\rightarrow \infty
\end{cases}\,.
\end{equation}
The corresponding local strain along the soliton can  be written as
\begin{align}
\left(\partial _{x} u_{x} (x), \partial_{x} u_{y} (x)  \right) =
 \left( f '(x) \cos (\theta_B) ,f ' (x) \sin(\theta_B)\right)\,.
\end{align}
Using the fact that the elastic energy of a two dimensional solid is given by
\begin{multline}
E_\text{elas} = \frac{\lambda}{2} \int_{-\infty}^{\infty} (\partial _{x} {u}_{x}(x)) ^ {2} dx\\
 +\mu \int_{-\infty}^{\infty} \left[ (\partial _{x} {u}_ {x} (x)) ^ {2} + (\partial _{x} {u}_{y} (x))^{2} \right] ^2 dx.
\end{multline}
We find that the soliton's energy per unit length is given by
\begin{multline}
{E}_\text{elas} = \left[ \lambda \frac{\cos ^ {2}(\theta_B)}{2} + \mu {\cos} ^ {2} (\theta_B) + \mu \frac{{\sin} ^ {2} (\theta_B)}{2} \right]
\\\times\int_{-\infty}^{\infty} \left[f'(x) \right] ^ {2} dx\,.
\end{multline}
This is the result shown in Fig.~2d of the main text.

\bibliographystyle{nature}
\bibliography{Biblio_swirl}

\section*{Acknowledgements}
VTR, GTdL and FM acknowledge the support from ANR Flatmoi grant (ANR-21-CE30-0029). VTR and FM acknowledge the support from CEA PTC-instrum Dform. NRW acknowledges the support from the Science and Technology funding Council ST/V001116/1.  

\section*{Author contribution}
VTR, CC and FM conceived the experiments. CP and AM fabricated the samples. FM, VTR and CC performed and analyzed STM experiments. DD and HO performed TEM measurements. NRW and FG produced the theoretical models of atomic relaxation and GTdL the calculations of electronic structure. FM and VTR wrote the manuscript with the input of all authors. VTR supervised the collaboration.  

\section*{Competing financial interest}
The authors declare no competing financial interests.

\section*{Data availability}
The datasets generated during and/or analysed during the current study are available from the corresponding author on reasonable request.

\clearpage

\setcounter{figure}{0}
\renewcommand{\figurename}{\textbf{Supplementary Fig.}}   

\end{document}